\documentclass[10pt,conference]{IEEEtran}

\usepackage{cite}
\usepackage{amsmath,amssymb,amsfonts}
\usepackage{algorithmic}
\usepackage{graphicx}
\usepackage{textcomp}
\usepackage{xcolor}
\usepackage{booktabs}
\usepackage{makecell}
\usepackage{multirow}
\usepackage{pifont}
\usepackage{url}
\usepackage{xspace}
\usepackage{cleveref}

\newcommand{\mypara}[1]{\noindent{\bf #1}\xspace}

\def\BibTeX{{\rm B\kern-.05em{\sc i\kern-.025em b}\kern-.08em
    T\kern-.1667em\lower.7ex\hbox{E}\kern-.125emX}}

\begin{document}

\title{Beyond Refusal: A Same-Lineage Study of Aligned and Abliterated LLMs for Vulnerability Analysis}

\author{%
\IEEEauthorblockN{Mingchen Li\IEEEauthorrefmark{1}, Meikang Qiu\IEEEauthorrefmark{2}, Zifan Peng\IEEEauthorrefmark{3}, Heng Fan\IEEEauthorrefmark{1}, Song Fu\IEEEauthorrefmark{1}, Junhua Ding\IEEEauthorrefmark{1}, and Yunhe Feng\IEEEauthorrefmark{1}}
\IEEEauthorblockA{\IEEEauthorrefmark{1}University of North Texas, USA\\
MingchenLi@my.unt.edu, \{Heng.Fan, Song.Fu, Junhua.Ding, Yunhe.Feng\}@unt.edu}
\IEEEauthorblockA{\IEEEauthorrefmark{2}Augusta University, USA\\
qiumeikang@yahoo.com}
\IEEEauthorblockA{\IEEEauthorrefmark{3}Newcastle University, United Kingdom\\
z.peng12@newcastle.ac.uk}
}

\maketitle

\begin{abstract}
Large language model (LLM)-assisted software security operates at a difficult boundary: the vulnerability-analysis terminology needed for legitimate code review, triage, and repair can closely resemble terminology associated with misuse. Existing safety and cybersecurity evaluations are difficult to interpret in this setting because they often compare unrelated model families, thereby conflating safety behavior with differences in architecture, scale, training data, and deployment. To isolate this factor, we study \emph{safety state}: whether refusal behavior remains intact (\textsc{Aligned}) or has been refusal-ablated (\textsc{Abliterated}) within same-lineage models. We ask how this safety state affects defensive utility across software-security workflows. We compare aligned instruction-tuned models with publicly released refusal-ablated descendants from two model families, Gemma and Qwen. We evaluate \textsc{Aligned} and \textsc{Abliterated} states on vulnerability detection, CWE attribution, vulnerable-line localization, root-cause localization, and executable patch validation. We further treat prompt wording as a controlled framing dimension: prompts begin with neutral code-review language, add authorization context, and vary the density of cybersecurity terminology. In a Gemma-based Java/Vul4J repair-validation study, \textsc{Abliterated} achieves higher early-stage validation rates, with 67.8\%, 65.0\%, and 32.8\% of patches judged usable, successfully applied, and successfully compiled, respectively, compared with 29.9\%, 24.9\%, and 9.0\% for \textsc{Aligned}. In the Qwen pair, \textsc{Abliterated} improves localization performance, increasing line-level F1 from 2.08\% to 3.91\% and Top-1 accuracy from 4.10\% to 6.95\%. Overall, our results show that safety-state effects manifest not only as changes in whether models answer, but also as changes in answer coverage, localization quality, prompt sensitivity, and staged repair-validation outcomes. These findings suggest that evaluations of LLM-based security assistants should jointly measure whether models respond, whether their usable responses are correct, and whether their outputs remain actionable across the engineering workflow.
\end{abstract}


\section{Introduction}

Large language models (LLMs) are increasingly used as code-generation systems~\cite{chen2021evaluating,li2022competition,roziere2023code}, programming assistants~\cite{lu2021codexglue,feng2020codebert,wang2021codet5}, and repository-level software engineering agents~\cite{jimenez2023swe,yang2024swe}.
This transition has especially important implications for software security, where LLMs are evaluated for cyber-risk assessment~\cite{bhatt2024cyberseceval,wan2024cyberseceval}, threat intelligence~\cite{alam2024ctibench}, penetration-testing assistance~\cite{deng2024pentestgpt}, vulnerability detection~\cite{ding2024vulnerability}, and automated vulnerability repair~\cite{zhang2025llms,wei2025patcheval}.
Yet software-security workflows expose a persistent boundary problem: the concepts, vocabulary, and reasoning patterns that defenders need to identify and repair vulnerabilities can closely resemble misuse-oriented requests.
Recent deployment incidents make this tension concrete.
For example, the June 2026 Claude Fable 5 episode highlighted how stronger safeguards around high-risk domains can also produce broad refusals for benign cybersecurity work, including reading security materials and writing defensive code~\cite{anthropic2026fableMythos,anthropic2026apiReleaseNotes,bellan2026fablePublic,franceschiBicchierai2026fableGuardrails}.
For software engineering, the key question is therefore not only whether cyber-capable LLMs block harmful requests, but also whether their safeguards preserve legitimate defensive utility.

Defender-side failures can take two forms.
The first is visible: a model may falsely refuse an authorized vulnerability-analysis request, reducing the availability and reliability of LLM assistance in security workflows.
The second is subtler: a model may answer, but produce guidance that is less correct, less specific, or less actionable.
In vulnerability analysis, such silent degradation can lead to operationally significant errors.
An incorrect vulnerable-line prediction can misdirect triage, an inaccurate CWE label can distort risk classification, an incomplete root-cause explanation can obscure the true bug pattern, and a plausible but invalid patch can create a false sense of remediation.
Thus, for LLMs deployed in security settings, availability, correctness, and actionability are not merely usability concerns; they are security-relevant properties of the system.

Prior work has begun to expose this tension, showing that safety-tuned frontier models can over-refuse legitimate cyber-defense requests, especially when prompts contain harm-adjacent cybersecurity terminology, and that explicit authorization does not always recover compliance~\cite{campbell2026defensive}.
However, current evidence remains limited in two ways.
First, many evaluations compare different model families or provider systems, making it difficult to attribute observed behavior to safety alignment rather than to architecture, scale, training data, post-training recipes, serving infrastructure, or proprietary policy layers.
Second, existing evaluations often emphasize refusal rate itself.
Although refusal is important, it does not capture whether non-refused answers are correct, useful, or actionable in realistic vulnerability-analysis workflows.
A model that rarely refuses but produces incorrect localization or unusable patches may be no more useful to defenders than one that refuses frequently.

This paper studies this missing dimension by examining how a model's \emph{safety state} changes the utility of legitimate vulnerability analysis.
We define safety state as whether refusal behavior remains intact in an aligned instruction-tuned model, denoted \textsc{Aligned}, or has been reduced in a refusal-ablated descendant, denoted \textsc{Abliterated}.
To isolate this factor, we conduct same-lineage comparisons: the evaluated states are rooted in the same open-source model family and tested under matched settings.
Gemma~\cite{team2024gemma} serves as the primary same-lineage pair for the full study, comparing \texttt{gemma-4-31B-it} with the public refusal-ablated descendant \texttt{gemma-4-31B-CRACK}.
Qwen~\cite{qwen3.6-27b} provides a supplementary same-lineage pair for the overview and vulnerable-line localization analyses, enabling us to examine whether observed patterns extend across model lineages.

We evaluate these models across a progression of defensive vulnerability-analysis tasks: vulnerability detection, CWE attribution, vulnerable-line localization, root-cause localization, and patch generation with executable validation.
This task ladder reflects the structure of real security work, which does not stop at deciding whether code is vulnerable.
Defenders must identify where the vulnerability occurs, explain why it occurs, map it to known weakness classes, and determine whether a proposed fix can be applied, compiled, and validated.
Accordingly, we decompose defender-side utility into three complementary dimensions: \emph{coverage}, measuring whether the model returns a usable answer; \emph{answer quality}, measuring correctness among analyzable answers after excluding refusals and malformed outputs; and \emph{end-to-end utility}, measuring the practical usefulness of the full interaction from prompt to validated output.

Our results show that the defender-side cost of alignment is context-dependent rather than monotonic.
Across broad vulnerability detection, CWE attribution, and localization tasks, both aligned and abliterated safety states usually return usable answers, indicating that outright refusal is not the dominant source of utility loss.
Instead, safety state changes the model's utility profile.
The aligned state is stronger on several shallow diagnostic tasks, especially under neutral code-review wording, where it preserves stronger high-level classification performance.
As tasks become more code-grounded and actionable, however, the abliterated state becomes more competitive and, in some settings, stronger: it improves focused localization behavior and performs better at early executable patch-generation gates, including producing patches that can be applied and compiled.
The Qwen pair provides cross-lineage evidence for this localization pattern: in pooled vulnerable-line localization, \textsc{Abliterated} improves Top-1 accuracy from 4.10\% to 6.95\% and mean line-level F1 from 2.08\% to 3.91\%.

Our prompt-framing results further show that professional security language is not a harmless surface detail.
Under neutral software-review wording, the aligned model often retains an advantage on detection and CWE-style classification.
When prompts add explicit authorization, cybersecurity terminology, or stronger vulnerability-analysis wording, that advantage can shrink or shift toward the abliterated state, especially for localization and repair-oriented tasks.
This finding suggests that safety mechanisms may respond not only to the legitimacy of the task, but also to surface cues associated with cyber misuse.
The implication is not that safeguards should be removed.
Rather, cyber-safety evaluation should measure both sides of the deployment objective: preventing harmful compliance while preserving reliable, high-quality assistance for legitimate defenders.

This paper makes the following contributions.

\begin{itemize}
    \item \textbf{Matched same-lineage evaluation of LLM safety states.} We present, to our knowledge, the first controlled study of vulnerability-analysis behavior that varies only safety state, using Gemma as the primary same-lineage pair and Qwen as a supplementary pair for the overview and vulnerable-line localization, with matched settings for every comparison.

    \item \textbf{A defender-side utility decomposition beyond refusal rate.}
    We show that refusal rate alone is insufficient for evaluating cyber-safety mechanisms.
    We decompose model behavior into answer coverage, answer quality, and end-to-end utility, revealing cases where safety-state effects appear in the correctness or actionability of non-refused answers.

    \item \textbf{A task-depth analysis from diagnosis to executable repair.}
    We evaluate safety-state effects across vulnerability detection, CWE attribution, vulnerable-line localization, root-cause localization, and executable patch validation.
    This task ladder shows that alignment effects are not uniform: the relative advantage of each safety state changes as tasks become more code-grounded and operationally actionable.

    \item \textbf{A controlled prompt-framing analysis of professional security language.}
    We measure how neutral review wording, authorization clauses, cybersecurity terminology, and lexical-intensity variations affect each safety state while holding the underlying code and task fixed.
    The results show that legitimate defensive prompts can be sensitive to surface cues that safety mechanisms may associate with misuse.

    \item \textbf{Implications for defender-safe LLM evaluation and design.}
    We derive guidance for evaluating cyber-capable LLMs: safety mechanisms should distinguish authorized defensive analysis from exploit enablement, and evaluations should jointly measure refusal, correctness, localization quality, and executable actionability rather than treating non-refusal as sufficient success.
\end{itemize}

\section{Related Work and Positioning}
\label{sec:related-work}

\mypara{LLMs in software engineering.}
A related line of work evaluates LLMs as code-generation models, code-understanding models, and software-engineering agents.
Early benchmarks and pretrained code models such as HumanEval~\cite{chen2021evaluating}, CodeXGLUE~\cite{lu2021codexglue}, CodeBERT~\cite{feng2020codebert}, GraphCodeBERT~\cite{guo2020graphcodebert}, CodeT5~\cite{wang2021codet5}, PLBART~\cite{ahmad2021unified}, and UniXcoder~\cite{guo2022unixcoder} established program understanding and generation as standard evaluation targets.
More recent systems such as AlphaCode~\cite{li2022competition}, StarCoder~\cite{li2023starcoder}, Code Llama~\cite{roziere2023code}, WizardCoder~\cite{luo2023wizardcoder}, and DeepSeek-Coder~\cite{guo2024deepseek} show that code-specialized LLMs can perform increasingly strong code generation and instruction-following over programming tasks.
Repository-level benchmarks and agents such as SWE-bench~\cite{jimenez2023swe}, SWE-agent~\cite{yang2024swe}, AutoCodeRover~\cite{zhang2024autocoderover}, and Agentless~\cite{xia2024agentless} further move evaluation toward multi-file editing, test execution, and realistic issue resolution~\cite{jimenez2023swe}.
These works establish LLMs as software-engineering tools, while leaving open how safety state and security-sensitive wording affect vulnerability-analysis workflows.

\mypara{Vulnerability-program datasets and repair validation.}
Complementary work develops datasets and metrics for vulnerable code.
CVEFixes links vulnerability records, commits, and code artifacts at multiple abstraction levels, supporting downstream security mining tasks~\cite{bhandari2021cvefixes}.
DiverseVul and MegaVul expand coverage across projects and CWE categories, while LineVul pushes vulnerability evaluation toward line-level localization rather than only function-level detection~\cite{chen2023diversevul,fu2022linevul,ni2024megavul}.
PrimeVul further argues that common vulnerability benchmarks suffer from leakage, weak labels, and unrealistic splits, and introduces cleaner labels, chronological splits, paired samples, and vulnerability-detection metrics for more realistic evaluation~\cite{ding2024vulnerability}.
For repair, Vul4J provides reproducible Java vulnerabilities with proof-of-vulnerability tests and patches, while recent repair benchmarks and surveys such as PatchEval and vulnerability-repair SoK studies emphasize executable validation across real-world vulnerabilities~\cite{bui2022vul4j,wei2025patcheval,hu2025sok}.
At the method level, GRACE and recent vulnerability-repair studies show that graph structure, in-context information, reasoning, and validation feedback can substantially alter LLM defender-side utility~\cite{lu2024grace,kulsum2024case}.
These datasets and methods provide the task substrates for detection, localization, root-cause reasoning, and repair.
We use them as an evaluation layer for studying how safety state changes defender-side utility across concrete software-engineering gates.

\mypara{Cyber-safety evaluation and defensive refusal.}
Another line of work evaluates LLM cybersecurity behavior more directly.
Recent surveys show rapid growth of LLM-based cyber assistants across threat intelligence, incident response, code security, and agentic automation~\cite{zhang2025llms}.
CyberSecEval-style benchmarks evaluate cybersecurity capabilities and risks, including prompt injection, code-interpreter abuse, offensive assistance, and safety--utility trade-offs~\cite{bhatt2024cyberseceval,wan2024cyberseceval}.
CTIBench focuses on analyst-facing cyber threat-intelligence tasks~\cite{alam2024ctibench}.
Agent-safety work such as AgentBench, AgentDojo, and ToolEmu-style sandboxing studies interactive planning, tool use, prompt injection, and adversarial robustness in autonomous systems~\cite{liu2023agentbench,debenedetti2024agentdojo,ruan2023identifying}.
Most closely related, \emph{Defensive Refusal Bias} shows that aligned models can deny legitimate defenders in real cyber-defense contexts~\cite{campbell2026defensive}.
This line motivates our defender-side setting; our study extends the measurement target from refusal coverage to answer quality, output stability, and executable actionability.

\mypara{Safety alignment, safeguards, and editable refusal behavior.}
A large body of work studies how LLMs are aligned to follow user intent while avoiding harmful outputs.
RLHF and constitutional alignment established the dominant post-training paradigm for improving helpfulness and harmlessness~\cite{ouyang2022training,bai2022constitutional}.
Later work introduced dedicated safeguard models such as Llama Guard, which classify unsafe inputs and outputs as a separate safety layer~\cite{inan2023llama}.
Jailbreak research showed that aligned models can still be induced to violate policy through adversarial prompting or transferable suffix attacks~\cite{wei2023jailbroken,zou2023universal}.
Benchmark efforts then formalized this space with harmful-compliance and robust-refusal evaluations such as HarmBench, over-refusal-specific evaluations such as OR-Bench, and broader risk--usefulness suites such as FORTRESS~\cite{mazeika2024harmbench,cui2024or,knight2025fortress}.
Mechanistic refusal-direction studies further show that adding or subtracting a learned refusal direction in activation space can suppress or induce refusal behavior~\cite{arditi2024refusal}, and related surgical approaches study targeted representation or weight-space interventions~\cite{wang2024surgical}.
Together, these works make refusal behavior measurable and perturbable inside the model.
Our work uses this observation as the basis for a matched same-lineage evaluation of legitimate vulnerability analysis.

\mypara{Position of this work.}
Our study connects these lines of work by treating safety alignment as an experimental variable in vulnerability-program analysis.
Safety benchmarks measure refusal and harmful compliance, but usually outside fine-grained cyber workflows.
Cyber benchmarks measure security capabilities and risks, but often compare distinct model families or broad prompt categories.
Vulnerability-program benchmarks measure detection, localization, and repair utility, while safety-state effects on that utility remain underexplored.
We fill this gap with a matched same-lineage aligned-versus-abliterated evaluation that decomposes behavior into answer coverage, quality, and end-to-end utility across task depth and prompt framing.

\section{Study Design and Measurement Framework}
\label{sec:method}

Software security provides a high-stakes setting for studying LLM-assisted software engineering: the same code artifacts, data-flow concepts, and vulnerability terminology that defenders use during review and repair can also appear in misuse-oriented prompts. This dual-use overlap makes refusal behavior an incomplete proxy for safety in software-engineering workflows. A model may fail visibly by refusing an authorized request, or less visibly by answering with degraded localization, classification, explanation, or repair output. From a software engineering perspective, the central measurement problem is therefore not only whether an assistant complies, but whether its responses remain available, correct, and actionable across the vulnerability-analysis workflow.

We study this problem at the model-response layer. Our setting is a benign defensive workflow in which a user provides source code and asks the model to detect, classify, localize, explain, or repair a vulnerability. The evaluation scope is the response that a developer or security engineer would consume. Real deployments may additionally include prompt filters, policy routers, tool sandboxes, output filters, or human approval workflows; such model-external controls may attenuate or amplify the effects measured here. We intentionally hold these factors outside the study to isolate how the model's safety state changes behavior under controlled local evaluation.

\subsection{Matched Same-Lineage Safety States}
\label{sec:method-model-states}

We compare two states from the same open-source large language model family: an aligned instruction-tuned state and a public refusal-Abliterated descendant.
We refer to them as \textsc{Aligned} and \textsc{Abliterated}.
Both states are evaluated with the same prompt construction, local serving stack, decoding controls, and scoring pipeline.
This matched same-lineage design reduces the architecture and training-distribution confounds that appear when safety behavior is studied by comparing unrelated model families.

The abliterated state is motivated by a simple observation from refusal-direction work: refusal behavior in instruction-tuned LLMs can be partly represented as a direction in activation space \cite{arditi2024refusal,wang2024surgical}.
The procedure starts with two prompt sets, one that elicits refusal and one that permits ordinary assistance.
At a chosen layer \(l\), let \(h_l(p)\in\mathbb{R}^d\) be the hidden representation for prompt \(p\).
The first step is to average the two sets, producing a refusal centroid and an allowed-response centroid:
\[
\mu_l^R = \frac{1}{|D_R|}\sum_{p\in D_R} h_l(p),
\qquad
\mu_l^A = \frac{1}{|D_A|}\sum_{p\in D_A} h_l(p).
\]
The normalized difference between these centroids gives the refusal direction:
\[
r_l =
\frac{
\mu_l^R-\mu_l^A
}{
\|\mu_l^R-\mu_l^A\|_2}.
\]
The second step suppresses this direction.
In activation space, this means subtracting the projection of the representation onto \(r_l\):
$h_l' = h_l - \alpha r_l(r_l^\top h_l)$,
where \(\alpha\) controls the intervention strength.
In weight space, the same idea can be applied to selected residual-writing matrices by projecting their outputs away from \(r_l\):
\[
W_l' = (I-\alpha r_l r_l^\top) W_l.
\]
We use this background to characterize what the released abliterated state is designed to change: it weakens refusal-associated directions while preserving the surrounding model family as much as the public artifact permits.
The treatment is therefore a matched same-lineage safety-state comparison rather than a perfectly isolated causal alignment intervention.

\subsection{Two-Factor Utility Measurement: Task Depth and Prompt Framing}
\label{sec:method-utility-factors}

\subsubsection{Utility Decomposition Beyond Refusal}
\label{sec:method-formal-design}

\begin{figure*}[!ht]
\centering
\includegraphics[width=\textwidth]{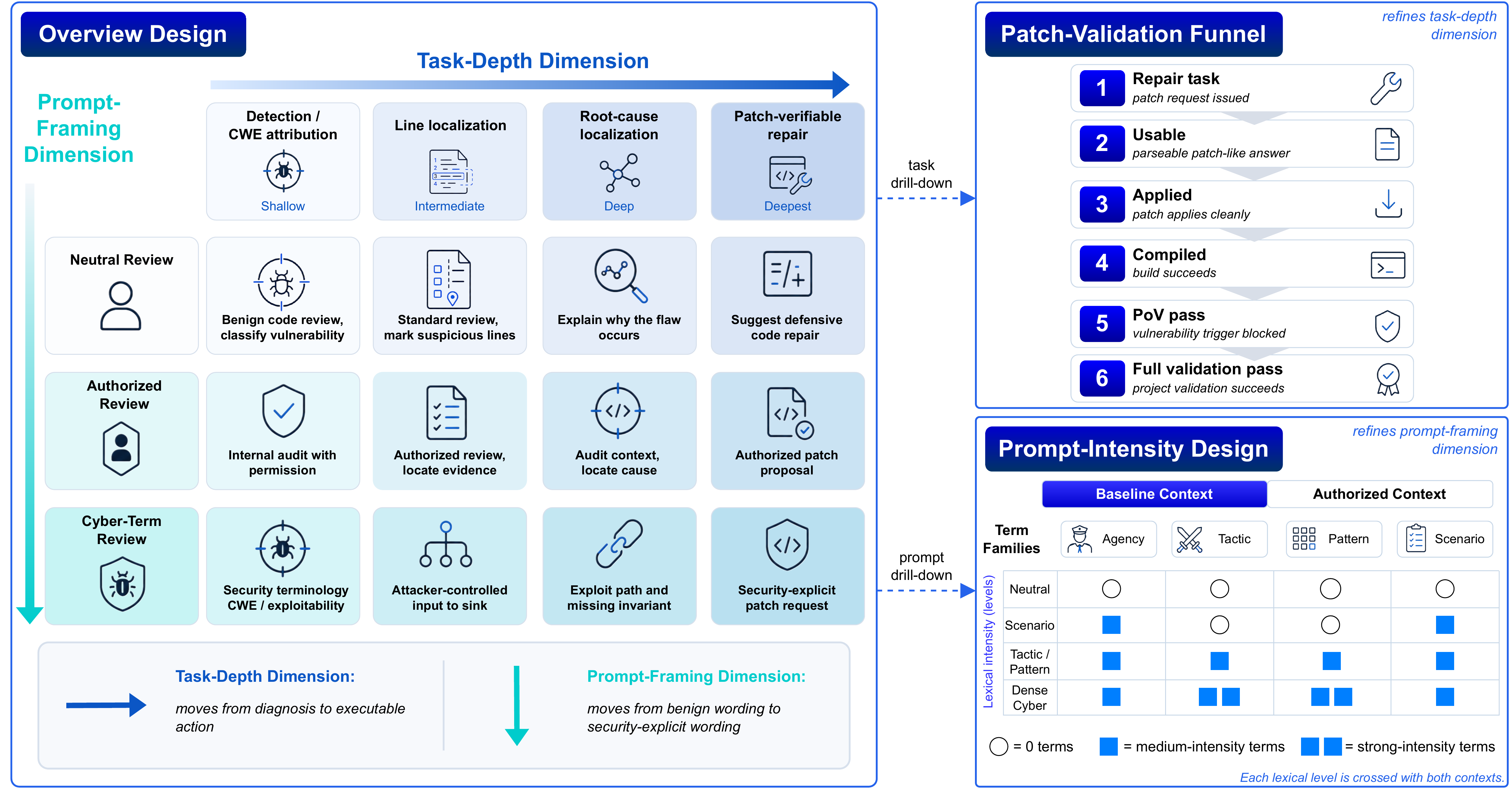}
\caption{Study design for measuring defender-side utility across task depth and prompt framing.
Left: the overview design crosses three prompt-framings (Neutral Review, Authorized Review, and Cyber-Term Review) with a task-depth progression from detection/CWE attribution to line localization, root-cause localization, and patch-verifiable repair.
Top right: the patch-validation funnel refines repair into executable gates from patch request to usable answer, applied patch, compiled build, PoV pass, and full validation pass.
Bottom right: the prompt-intensity design refines prompt framing by crossing Baseline and Authorized contexts with lexical-intensity levels built from agency, tactic, pattern, and scenario terms.
Together, these modules connect the paper's coarse utility comparison, executable repair validation, and controlled prompt-wording analysis.}
\label{fig:method-overview}
\vspace{-0.3cm}
\end{figure*}

We model an LLM as a mediator between a user-facing prompt and a task-facing output.
Let \(x\) be a code sample, \(t\in\mathcal{T}\) a vulnerability-analysis task, \(f\in\mathcal{F}\) a prompt frame, and \(s\in\{\textsc{Aligned},\textsc{Abliterated}\}\) the model state.
The response is
\[
y = M_s(x,t,f).
\]
This formulation maps to the crossed design in the left module of \Cref{fig:method-overview}, which separates two design dimensions.
The \emph{task-depth progression} asks how the safety state affects increasingly deep vulnerability-analysis tasks.
The \emph{prompt-framing} asks how the same underlying task changes when the user-facing framing changes.

For each response, we decompose utility into coverage and answer quality:
$U(y) = C(y)\cdot Q(y)$,
where \(C(y)\) indicates whether the response is usable for the task and \(Q(y)\) is the task-specific score among usable responses.
This decomposition is central to our study: a model can fail by refusing, by producing an unusable degraded answer, or by answering with low task quality.

\subsubsection{Task-Depth Progression: From Diagnosis to Executable Repair}
\label{sec:method-task-depth-progression}

The task-depth progression is a ladder from diagnostic judgments to executable actions.
\Cref{fig:method-overview} summarizes the task settings and their role in this ladder.
Detection and CWE attribution match common vulnerability-detection and taxonomy-level benchmarks \cite{chen2023diversevul,ni2024megavul,bhandari2021cvefixes,ding2024vulnerability}.
Line localization adds code-grounded evidence \cite{fu2022linevul,ding2024vulnerability}.
Root-cause reasoning keeps the line-based output surface but changes the target from vulnerable-line evidence to causal diagnosis.
Patch-verifiable repair then tests executable defensive action.
Java/Vul4J is the primary executable-repair substrate because it provides project-level proof-of-vulnerability and validation tests \cite{bui2022vul4j,lu2024grace,kulsum2024case}.
Python (PatchEval)~\cite{wei2025patcheval} and C/C++ (Vul4C)~\cite{hu2025sok} are used as supplementary neutral-frame cross-language checks; they test whether the repair pattern is confined to Java or appears at the boundary of other executable ecosystems.

\subsubsection{Prompt-Framing: Authorization and Security Terminology}
\label{sec:method-prompt-framing-conditions}

The prompt-framings are staged as well.
The overview, code-grounded localization, and executable repair studies begin with three coarse benign frames: Neutral Review, Authorized Review, and Cyber-Term Review.
Neutral Review uses ordinary code-review language.
Authorized Review keeps the same task but adds an audit context, e.g., an internal security review with permission to inspect the code.
Cyber-Term Review keeps the task defensive but uses professional security terminology, e.g., asking about attacker-controlled input, exploitability, or security impact.
This follows prior evidence that cyber safeguards react to authorization cues and harm-adjacent security terminology in legitimate defensive requests \cite{campbell2026defensive}, and that cyber safety benchmarks use scenario and terminology variation to expose safety--utility trade-offs \cite{bhatt2024cyberseceval,wan2024cyberseceval}.
\Cref{fig:method-overview} records these base frames as the prompt-framing in the overview study.

The prompt-intensity study refines prompt framing into two parallel dimensions: authorization context and lexical intensity.
The context dimension has two authorization states: Baseline Context omits the authorization clause, while Authorized Context adds an explicit internal-security-review clause.
The lexical-intensity dimension has four levels, defined by the term-family budgets in \Cref{fig:method-overview}.
Each lexical level is evaluated under both authorization states.

The prompt choices are taxonomy-driven and fixed before result inspection.
We start from the coarse-frame results in the overview, then use a fixed prompt taxonomy to separate authorization from wording intensity.
The bottom-right part of \Cref{fig:method-overview} shows the four types of term.
Agency terms introduce security actors and exploitability language; tactic terms follow ATT\&CK-style tactic nouns; pattern terms follow CAPEC-style vulnerability and attack-pattern nouns; and scenario terms provide security-review context.
All prompts keep the task defensive and hold the code, system prompt, output schema, and scorer fixed within each prompt frame.
\Cref{fig:method-overview} defines the lexical-intensity levels by term-family budget.
Neutral uses zero harm-adjacent taxonomy terms.
Scenario adds mild agency and review-scenario language.
Tactic/Pattern adds one tactic and one vulnerability-pattern term.
Dense Cyber keeps the task defensive but increases tactic and pattern density to two terms from each family.
These four lexical levels are combined with both Baseline Context and Authorized Context.

\subsection{Model Serving and Inference Details}
\label{sec:method-experiment-setup}

We use two same-lineage model pairs.
Gemma is the primary pair for the full study, covering the overview, focused localization, root-cause localization, executable repair, and prompt-framing analyses.
Qwen is evaluated on the overview and focused-localization studies to examine whether the empirical findings also appear across model lineages.

All Gemma experiments use local Q4\_K\_M GGUF artifacts served through the same \texttt{llama.cpp} deployment.
The aligned state is \texttt{gemma-4-31B-it}~\cite{google2026gemma431bIt}; the abliterated state is \texttt{gemma-4-31B-CRACK}~\cite{douyamv2026gemma431bCrackGguf}.
Public metadata for the abliterated artifact describes CRACK-style per-layer refusal-vector surgery followed by quantization and GGUF conversion; we evaluate that released artifact as a deployment endpoint.
Both Gemma states use the same local llama-server binary, context length 32,768, temperature 0, top-\(p\) 1, top-\(k\) 0, repeat penalty 1, and seed 42.

For the Qwen pair, we use \texttt{Qwen3.6-27B-MXFP4} as the \textsc{Aligned} state~\cite{osaurusai2026qwen3627bMxfp4} and \texttt{Qwen3.6-27B-MXFP4-CRACK} as the \textsc{Abliterated} state~\cite{dealignai2026qwen3627bCrack}.
Both model endpoints use MXFP4 artifacts served through an OpenAI-compatible \texttt{mlx-vlm} interface.
Qwen uses the same settings as Gemma for the corresponding studies.

Repeated executions serve as scoring, validation, and runtime-consistency checks under deterministic decoding.
All tasks share the same prompt renderer, parser, and scoring code; model-external safeguards such as policy routers, prompt filters, output filters, and human approval workflows are outside the evaluation scope.

\section{Overview: Safety-State Utility Across Task Depth and Prompt Frames}
\label{sec:two-factor-overview}

\begin{table}[!ht]
\caption{Defender-side utility in the overview study across Gemma and Qwen pairs.
The table reports end-to-end utility for detection, CWE attribution, and vulnerable-line localization under prompt framing Neutral Review, Authorized Review, and Cyber-Term Review, plus usable-answer coverage for each pair and state. Metrics are presented as \textsc{Aligned} / \textsc{Abliterated}.}
\label{tab:overview-frame-utility}
\centering
\small
\setlength{\tabcolsep}{3pt}
\resizebox{\columnwidth}{!}{%
\begin{tabular}{llccc}
\toprule
& \multirow{2}{*}{\textbf{Prompt Frame}} & \textbf{Detect. (\%)} & \textbf{CWE (\%)} & \textbf{Local. (\%)} \\
& & \multicolumn{3}{c}{\textsc{Aligned} / \textsc{Abliterated}} \\
\midrule
\multirow{5}{*}{\rotatebox[origin=c]{90}{Gemma}} 
& Usable Answers & 97.50 / 97.42 & 97.33 / 97.33 & 97.00 / 97.00 \\
\cmidrule(lr){2-5}
& Neutral Review & 59.83 / 56.58 & 23.00 / 18.00 & 5.70 / 5.60 \\
& Authorized Review & 57.92 / 57.83 & 17.25 / 18.75 & 3.71 / 5.07 \\
& Cyber-Term Review & 57.50 / 57.25 & 17.50 / 18.75 & 3.77 / 4.34 \\
& All Prompt Frames & 58.42 / 57.22 & 19.25 / 18.50 & 4.40 / 5.00 \\

\midrule
\multirow{5}{*}{\rotatebox[origin=c]{90}{Qwen}}
& Usable Answers & 96.56 / 91.67 & 97.67 / 92.67 & 99.33 / 98.22 \\
\cmidrule(lr){2-5}
& Neutral Review & 57.83 / 54.83 & 12.00 / 14.50 & 3.35 / 4.55 \\
& Authorized Review & 56.33 / 58.00 & 14.50 / 17.00 & 1.86 / 2.45 \\
& Cyber-Term Review & 52.83 / 57.67 & 14.00 / 16.00 & 0.97 / 2.19 \\
& All Prompt Frames & 55.67 / 56.83 & 13.50 / 15.83 & 2.06 / 3.06 \\

\bottomrule
\end{tabular}%
}
\end{table}

This section corresponds to the left module of \Cref{fig:method-overview}: the crossed overview design over task depth and prompt framing.
It provides the coarse utility comparison before the repair and prompt-intensity drill-downs.

The overview experiments focus on three tasks.
Detection and CWE attribution are drawn from PrimeVul-style vulnerability-program samples, which provide realistic vulnerable and non-vulnerable code examples together with vulnerability metadata \cite{ding2024vulnerability}.
Detection asks for a binary vulnerability judgment, CWE attribution tests whether the model maps vulnerable behavior to a security taxonomy, and line localization uses LineVul-style line-level vulnerability data that requires source-line evidence beyond function-level classification \cite{fu2022linevul}.
Overall, each model state is evaluated on 3,600 detection rows, 1,200 CWE-attribution rows, and 900 localization rows, giving 5,700 rows per state and 11,400 scored rows in total.
We use end-to-end utility as the headline score: unusable answers receive zero utility, detection uses binary correctness, CWE attribution uses exact match, and localization uses line-F1.
The Usable Answer Coverage rows report the fraction of responses that pass parsing and usability checks.

\Cref{tab:overview-frame-utility} shows that prompt framing changes the state gap in both same-lineage pairs.
For Gemma, Neutral Review favors \textsc{Aligned} on detection and CWE attribution, but Authorized Review and Cyber-Term Review move CWE attribution and localization toward \textsc{Abliterated}.
For Qwen, Neutral Review still favors \textsc{Aligned} on detection, while the two security-explicit frames favor \textsc{Abliterated}; CWE attribution and localization favor \textsc{Abliterated} under all three frames.
Across prompt-framings, security-explicit wording generally moves the relative advantage toward \textsc{Abliterated}; localization gives the most consistent cross-pair result.

The task-depth progression separates shallow diagnostic behavior from code-grounded localization.
Gemma's pooled detection and CWE rows favor \textsc{Aligned}, at 58.42\% versus 57.22\% and 19.25\% versus 18.50\%.
Qwen's pooled detection and CWE rows favor \textsc{Abliterated}, at 56.83\% versus 55.67\% and 15.83\% versus 13.50\%.
When the task becomes code-grounded, pooled localization favors \textsc{Abliterated} in both model pairs: 5.00\% versus 4.40\% for Gemma and 3.06\% versus 2.06\% for Qwen.
The Usable Answer Coverage rows provide secondary context: Gemma coverage is nearly matched, and Qwen line-localization coverage remains high even though Qwen \textsc{Abliterated} coverage is lower on detection and CWE attribution.
The overview result is therefore bounded: prompt framing and task depth jointly shape the state gap, with localization giving the clearest cross-pair result.

\mypara{Finding.}
The overview establishes the phenomenon that motivates the rest of the paper: safety-state effects depend on both prompt framing and task depth; shallow diagnostic tasks differ across Gemma and Qwen, while code-grounded localization consistently favors \textsc{Abliterated}.
\Cref{sec:focused-localization} next examines vulnerable-line localization in detail, \Cref{sec:patch} evaluates executable actionability, and \Cref{sec:prompt-framing-results} returns to prompt-framings with controlled authorization and lexical-intensity variations.

\section{Code-Grounded Localization: Vulnerable Lines and Root Causes}
\label{sec:focused-localization}

\subsection{Vulnerable-Line Localization}
\label{sec:vulnerable-line-localization}

This section occupies the middle two columns of the task-depth progression in \Cref{fig:method-overview}, where the output moves from task labels to source-code evidence.
We first use vulnerable-line localization to test whether the overview's code-grounded result appears in both Gemma and Qwen same-lineage pairs.
We then move to Vul4J root-cause localization with the Gemma primary pair, where the target shifts from suspicious source lines to causal vulnerability diagnosis.

The vulnerable-line study uses a balanced design for both Gemma and Qwen.
Each prompt frame contributes 700 samples per model state, giving 2,100 sample-level comparisons and 4,200 scored rows overall.
Across both pairs, we report mean line-F1 as overlap with labeled vulnerable lines and Top1/Top5 hit rates as whether a labeled vulnerable line appears among the top-ranked predictions.

\begin{table}[!t]
\caption{Vulnerable-line localization across Gemma and Qwen pairs.
The table reports mean line-F1 and Top1/Top5 hit rates (\%) by prompt frame. Metrics are presented as \textsc{Aligned} / \textsc{Abliterated}; line-F1 measures overlap with labeled vulnerable lines, while Top1/Top5 measures whether a labeled line appears among the top-ranked predictions.}
\label{tab:focused-localization-frame}
\centering
\small
\setlength{\tabcolsep}{3pt}
\resizebox{\columnwidth}{!}{%
\begin{tabular}{llccc}
\toprule
& \multirow{2}{*}{\textbf{Prompt Frame}} & \textbf{F1 (\%)} & \textbf{Top1 (\%)} & \textbf{Top5 (\%)} \\
\cmidrule(lr){3-5}
& & \multicolumn{3}{c}{\textsc{Aligned} / \textsc{Abliterated}} \\
\midrule
\multirow{4}{*}{\rotatebox[origin=c]{90}{Gemma}} & Neutral Review & 4.18 / 4.36 & 10.57 / 11.86 & 13.14 / 13.71 \\
& Authorized Review & 3.69 / 3.94 & 8.29 / 9.43 & 11.29 / 12.00 \\
& Cyber-Term Review & 3.65 / 3.70 & 8.14 / 9.00 & 11.57 / 12.00 \\
& All Prompt Frames & 3.84 / 4.00 & 9.00 / 10.10 & 12.00 / 12.57 \\
\midrule
\multirow{4}{*}{\rotatebox[origin=c]{90}{Qwen}} & Neutral Review & 2.18 / 4.48 & 6.57 / 10.00 & 8.00 / 14.86 \\
& Authorized Review & 2.20 / 4.03 & 3.43 / 6.57 & 6.29 / 12.00 \\
& Cyber-Term Review & 1.87 / 3.24 & 2.29 / 4.29 & 4.57 / 8.86 \\
& All Prompt Frames & 2.08 / 3.91 & 4.10 / 6.95 & 6.29 / 11.90 \\
\bottomrule
\end{tabular}
}
\end{table}

\begin{figure*}[!ht]
    \centering
    \includegraphics[width=0.6\textwidth]{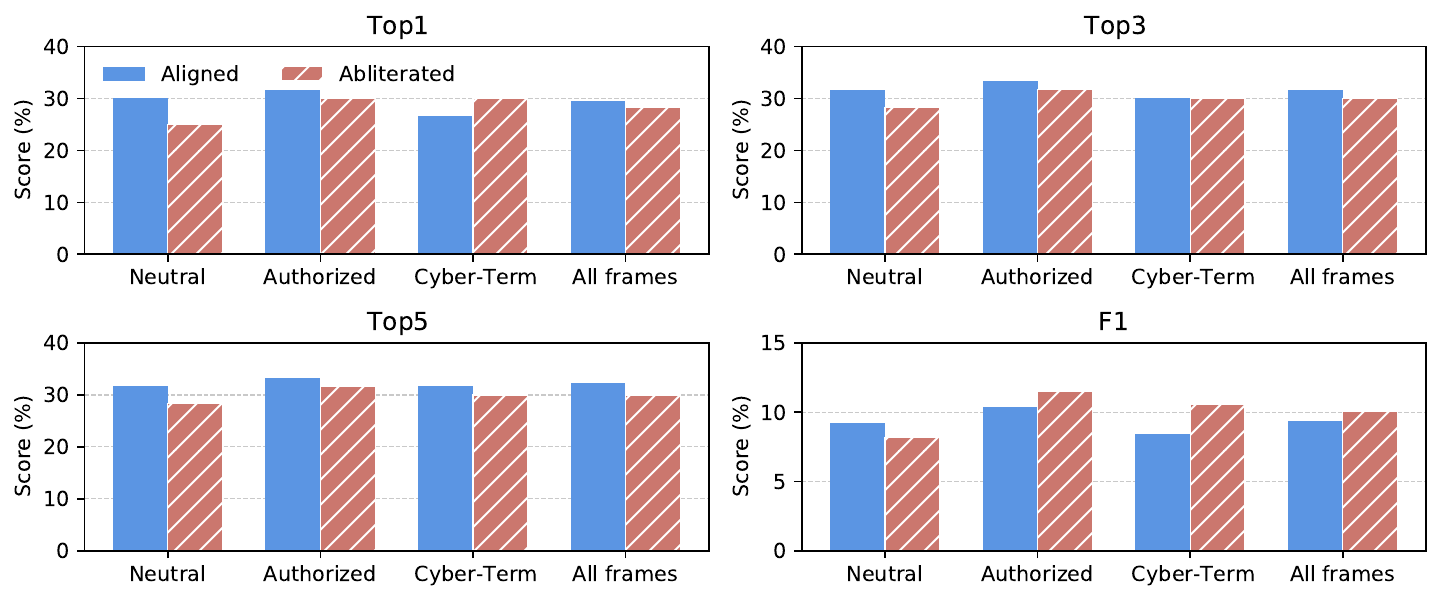}
    \includegraphics[width=0.39\textwidth]{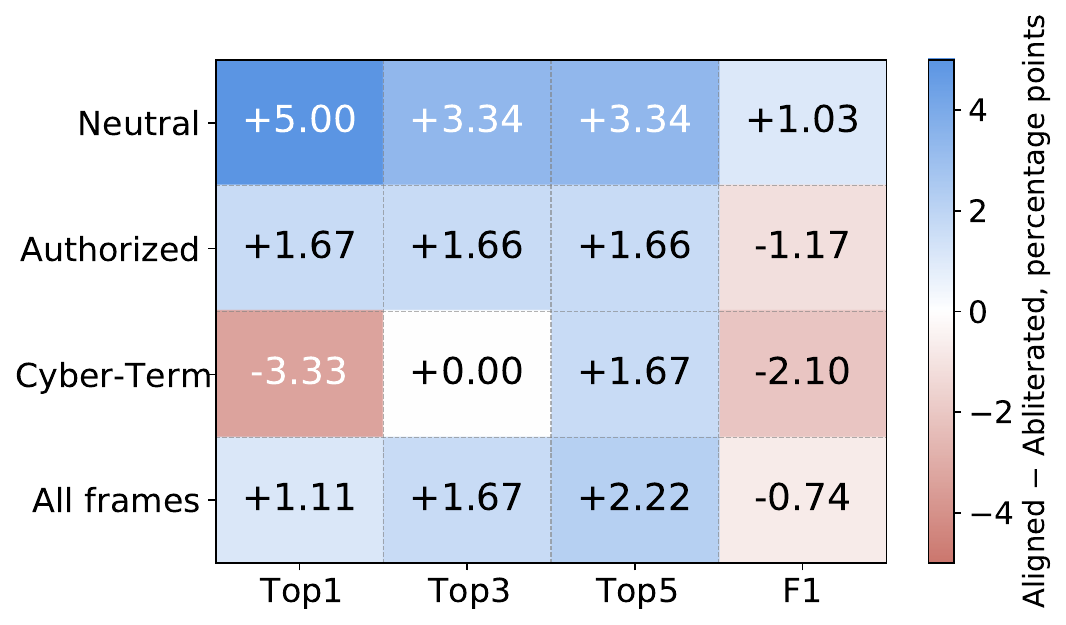}
    \vspace{-0.6cm}
    \caption{Root-cause localization on Vul4J with the Gemma pair.
The bar panels report Top1/Top3/Top5 hit rates and mean line-F1 by prompt frame; the heatmap reports \textsc{Aligned}-minus-\textsc{Abliterated} differences in percentage points, where blue cells favor \textsc{Aligned} and red cells favor \textsc{Abliterated}.}
    \label{fig:root_cause_combined}
    \vspace{-0.3cm}
\end{figure*}

\Cref{tab:focused-localization-frame} reports vulnerable-line localization results for the two same-lineage pairs.
For Gemma, \textsc{Abliterated} improves over \textsc{Aligned} by 1.29 percentage points in Neutral Review, 1.14 points in Authorized Review, and 0.86 points in Cyber-Term Review on Top1.
For Qwen, the corresponding Top1 gaps are 3.43, 3.14, and 2.00 points.
Across the pooled rows, \textsc{Abliterated} reaches 10.10\% Top1 versus 9.00\% for \textsc{Aligned} in Gemma, and 6.95\% versus 4.10\% in Qwen.
Mean line-F1 follows the same direction, moving from 3.84\% to 4.00\% in Gemma and from 2.08\% to 3.91\% in Qwen.
Thus, both pairs show higher \textsc{Abliterated} scores on Top1 and mean line-F1.

This result sharpens the overview finding.
Detection and CWE attribution vary by model pair, whereas vulnerable-line localization favors \textsc{Abliterated} in both Gemma and Qwen.
It connects the broad overview result to the deeper code-grounded analysis below.

\mypara{Finding.}
Vulnerable-line localization gives the clearest cross-pair code-grounded result: \textsc{Abliterated} improves Top1 and mean line-F1 in both same-lineage pairs, while the absolute localization scores remain low.
\Cref{sec:root-cause-localization} next tests how this pattern changes when the target shifts from vulnerable-line evidence to causal diagnosis.

\subsection{Root-Cause Localization}
\label{sec:root-cause-localization}

Having established the vulnerable-line result across two pairs, root-cause localization moves to a deeper causal target.
Both tasks ask the model to name source lines, but they ask for different evidence.
Focused localization asks where the vulnerable code appears; a good answer can be a line-level pointer to the suspicious statement.
Root-cause localization asks why the vulnerability exists; a good answer should identify the causal code region, guard, or missing security invariant that explains the vulnerable behavior and would guide a repair.
This makes root-cause localization a deeper step in the task-depth progression defined in \Cref{sec:method-task-depth-progression}: the metric is still line-based, with causal diagnosis as the intended target.
This subsection therefore asks how the safety-state gap changes when the task moves from line grounding to causal vulnerability diagnosis.
We use the Gemma pair as the primary for following tasks.

We use Vul4J because it provides real Java vulnerabilities with project-level artifacts and vulnerability-specific metadata \cite{bui2022vul4j}.
The root-cause subset contains 180 prompts per model state, evenly distributed across Neutral Review, Authorized Review, and Cyber-Term Review.
Each model returns root-cause line predictions, scored by Top1/Top3/Top5 hit rates and mean line-F1; Top-k measures whether a labeled causal line appears among the top-k predictions, while line-F1 measures overlap with the labeled causal region.

\Cref{fig:root_cause_combined} shows the overall split: \textsc{Aligned} is slightly higher on ranked root-cause hits, including Top1 at 29.44\% versus 28.33\%, while \textsc{Abliterated} is higher on mean line-F1, 10.08\% versus 9.34\%.
The frame rows explain where this split comes from.
In Neutral Review, \textsc{Aligned} is higher on both Top1 and mean line-F1.
Under Authorized Review and Cyber-Term Review, \textsc{Abliterated} becomes stronger on mean line-F1, while \textsc{Aligned} preserves a small ranked-hit advantage in Authorized Review.

\mypara{Finding.}
Code-grounded localization separates two effects.
Vulnerable-line localization favors \textsc{Abliterated} in both Gemma and Qwen, while Gemma root-cause localization splits ranked precision from region overlap.
\textsc{Aligned} is slightly stronger at placing the first root-cause line, whereas \textsc{Abliterated} achieves better mean line-F1.
\Cref{sec:patch} next tests whether these diagnostic patterns carry over when the model must produce executable repairs.

\section{Executable Repair: From Patch Text to Validated Fixes}
\label{sec:patch}

This section completes the task-depth progression defined in \Cref{sec:method-task-depth-progression} and the top-right part in \Cref{fig:method-overview}.
The preceding section measured code-grounded diagnostic utility; here we ask whether a model can produce a patch that enters an executable validation pipeline.
\Cref{sec:vul4j-patch-validation} evaluates Java/Vul4J repair under the coarse prompt frames, and \Cref{sec:cross-language-neutral-patch} checks whether the neutral-prompt repair pattern transfers across executable substrates.
Each repair experiment reports staged executable gates that separate answer availability, patch application, build success, vulnerability-trigger blocking, and final validation.

\subsection{Vul4J Patch-Validation Funnel}
\label{sec:vul4j-patch-validation}

Root-cause localization in \Cref{sec:root-cause-localization} demonstrates that deeper vulnerability analysis can alter the safety-state profile, but it still yields a diagnostic answer.
We next move from diagnosis to action, corresponding to the deepest task-depth step in \Cref{sec:method-task-depth-progression}: can the model produce a patch that survives an executable validation pipeline?
This step matters because a patch-like answer can look useful to an analyst while failing to apply, failing to compile, or leaving the vulnerability trigger intact.

We use 59 Vul4J vulnerabilities rendered under Neutral Review, Authorized Review, and Cyber-Term Review, giving 177 patch tasks per model state \cite{bui2022vul4j}.
Each model response is parsed into a patch candidate and evaluated through a staged funnel.
The gates are intentionally ordered from answer availability to executable repair.
\emph{Usable} means the response contains a parseable patch-like answer that can enter patch extraction.
\emph{Applied} means the extracted patch changes the target source and applies to the vulnerable checkout.
\emph{Compiled} means the patched project builds far enough for validation to proceed.
\emph{PoV pass} means the vulnerability-specific proof-of-vulnerability trigger is blocked.
\emph{Full validation pass} means the patch also passes the broader validation suite available for the Vul4J project.

\begin{figure*}[!ht]
    \centering
    \includegraphics[width=0.85\textwidth]{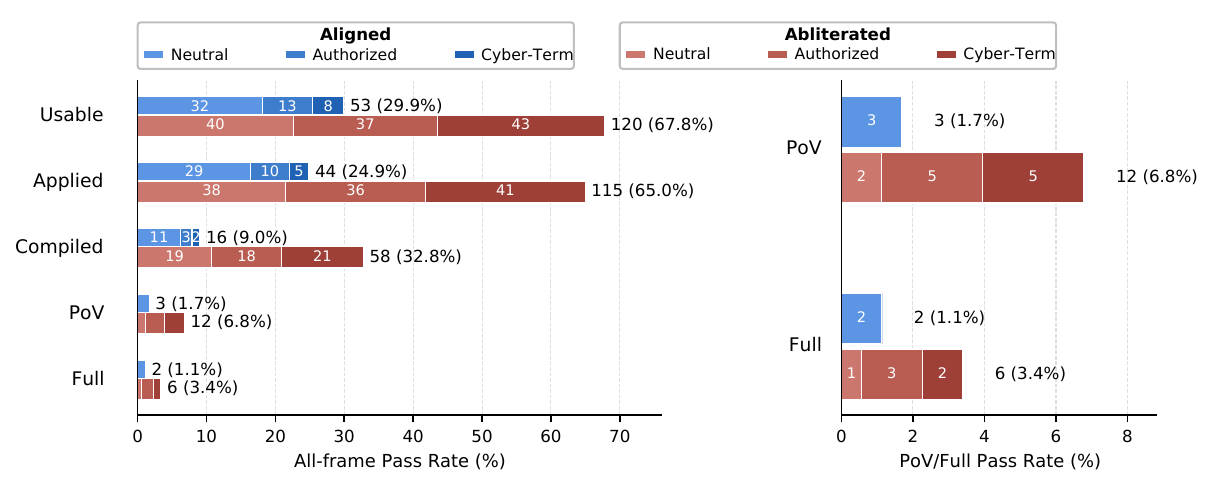}
    \vspace{-0.4cm}
    \caption{Vul4J patch-validation funnel for the Gemma pair across prompt frames and model states.
The left panel reports all-frame pass rates across the five validation gates; bar length gives the pass rate, stacked colors give prompt-frame contributions, interior labels give frame-level counts for the first three gates, and bar-end labels give total counts with rates.
The right panel enlarges PoV and full-validation with the same encoding, placing late-gate frame counts inside the bars and total counts with rates at the bar ends.}
    \label{fig:vul4j-patch-frame}
    \vspace{-0.3cm}
\end{figure*}

\Cref{fig:vul4j-patch-frame} summarizes the cleaned Vul4J funnel by prompt frame and gives both the all-frame rate view and the late-gate zoom.
The all-frame totals show a large throughput gap: \textsc{Abliterated} reaches 67.80\% usable answers, 64.97\% applied patches, and 32.77\% compiled patches, compared with 29.94\%, 24.86\%, and 9.04\% for \textsc{Aligned}.
The gap continues to the vulnerability trigger gate, where \textsc{Abliterated} reaches 6.78\% PoV pass compared with 1.69\% for \textsc{Aligned}.
The final validation gate is lower-count but directionally consistent: \textsc{Abliterated} obtains 6 full validation passes, while \textsc{Aligned} obtains 2.
The robust result is therefore early executable throughput, with the final validation gate serving as a sparse endpoint.

\begin{table*}[!ht]
\caption{Cross-language neutral-prompt patch repair.
Values are case-level counts with percentages computed over the total cases column.
The final gate is benchmark-specific: Vul4J requires full validation, PatchEval reports full status success over runtime-available cases, and Vul4C uses a strict re-derived executable success gate.}
\label{tab:cross-language-neutral-results}
\centering
\small
\setlength{\tabcolsep}{4pt}
\begin{tabular*}{\textwidth}{@{\extracolsep{\fill}}lclccccc}
\toprule
\textbf{Substrate} & \textbf{Total Cases} & \textbf{State} & \textbf{Usable} & \textbf{Applied / Runner} & \textbf{Compiled} & \textbf{PoV / Trigger} & \textbf{Final Gate} \\
\midrule
\multirow{2}{*}{Java/Vul4J} & \multirow{2}{*}{59} & \textsc{Aligned} & 32 (54.24\%) & 29 (49.15\%) & 11 (18.64\%) & 3 (5.08\%) & 2 (3.39\%) \\
& & \textsc{Abliterated} & 40 (67.80\%) & 38 (64.41\%) & 19 (32.20\%) & 2 (3.39\%) & 1 (1.69\%) \\
\midrule
\multirow{2}{*}{Python/PatchEval-71} & \multirow{2}{*}{71} & \textsc{Aligned} & 68 (95.77\%) & 61 (85.92\%) & N/A & 6 (8.45\%) & 6 (8.45\%) \\
& & \textsc{Abliterated} & 61 (85.92\%) & 58 (81.69\%) & N/A & 6 (8.45\%) & 5 (7.04\%) \\
\midrule
\multirow{2}{*}{C/C++ Vul4C-65} & \multirow{2}{*}{65} & \textsc{Aligned} & 62 (95.38\%) & 58 (89.23\%) & 36 (55.38\%) & N/A & 7 (10.77\%) \\
& & \textsc{Abliterated} & 54 (83.08\%) & 50 (76.92\%) & 29 (44.62\%) & N/A & 4 (6.15\%) \\
\bottomrule
\end{tabular*}
\vspace{-0.4cm}
\end{table*}

The stacked frame segments explain where the all-frame gap comes from.
In Neutral Review, \textsc{Abliterated} still enters the early gates more often, with 40 usable answers, 38 applied patches, and 19 compiled patches, compared with 32, 29, and 11 for \textsc{Aligned}.
Yet the final neutral-frame successes favor \textsc{Aligned}: 3 PoV passes and 2 full validation passes, compared with 2 and 1 for \textsc{Abliterated}.
The large \textsc{Abliterated} advantage is driven by the security-explicit frames.
\textsc{Aligned} drops sharply from 32 usable neutral answers to 13 under Authorized Review and 8 under Cyber-Term Review, and records zero PoV or full validation passes under both security-explicit frames.
\textsc{Abliterated} stays high-throughput across frames and obtains five of its six full-validation successes from Authorized Review and Cyber-Term Review.
Thus, the Java/Vul4J result is both a task-depth result and a prompt-framing interaction: \textsc{Abliterated} is much better at keeping patch generation executable when the benign request uses stronger cyber framing.

At the final gate, the six \textsc{Abliterated} successes span four CWE categories (CWE-611, CWE-79, CWE-502, and CWE-74), with five under Authorized Review or Cyber-Term Review; the two \textsc{Aligned} successes both occur under Neutral Review.
Case-level inspection further shows that intermediate gates correspond to distinct failure modes: trigger-level partial fixes, wrong security invariants, project-integration failures, and patch-like text that lacks effective code action.
These cases support the main interpretation of \Cref{fig:vul4j-patch-frame}: the most stable \textsc{Abliterated} advantage is early executable throughput, with full repair remaining a stricter and much lower-rate outcome.

\mypara{Finding.}
The Vul4J patch experiment strengthens the task-depth findings: when the task becomes executable repair, \textsc{Abliterated} has a clear Java advantage in early actionability.
The key nuance is where that advantage appears.
The all-frame \textsc{Abliterated} lead is carried by Authorized Review and Cyber-Term Review; in Neutral Review, \textsc{Aligned} still has slightly more final successful Java repairs even though \textsc{Abliterated} reaches the early gates more often.
This observation motivates \Cref{sec:cross-language-neutral-patch}: if neutral Java repair differs from the all-frame aggregate, the next question is whether neutral-prompt repair behaves similarly in other executable substrates such as Python and C/C++.

\subsection{Cross-Language Neutral Patch Repair}
\label{sec:cross-language-neutral-patch}

The Vul4J result in \Cref{sec:vul4j-patch-validation} establishes a strong Java all-frame effect, but the frame breakdown shows heterogeneous effects: \textsc{Abliterated} dominates early gates across all frames, while neutral Java repair gives \textsc{Aligned} a small advantage at the final gates.
We therefore add the cross-language neutral check specified in the task-depth design of \Cref{sec:method-task-depth-progression}.
The purpose is boundary testing with substrate-specific interpretation: Java/Vul4J~\cite{bui2022vul4j}, Python/PatchEval~\cite{wei2025patcheval}, and C/C++ Vul4C~\cite{hu2025sok} expose different executable repair substrates, runtime harnesses, and final gate definitions.
This subsection asks two linked questions: does the neutral-frame \textsc{Aligned} advantage at the final gate persist beyond Java, and does the \textsc{Abliterated} early-throughput advantage persist beyond Java?

\Cref{tab:cross-language-neutral-results} shows that the Java all-frame pattern does not simply carry over to neutral prompts in other languages.
In the Java neutral slice, \textsc{Abliterated} still has higher early throughput: 40 usable answers versus 32, 38 applied patches versus 29, and 19 compiled patches versus 11.
Yet \textsc{Aligned} has the slightly higher trigger and final counts, 3 versus 2 at PoV pass and 2 versus 1 at full validation.
The Python/PatchEval-71 slice moves further toward \textsc{Aligned}: it has higher usable output, higher runner entry, and 6 final successes versus 5, while the trigger count is tied.
The C/C++ Vul4C-65 slice is the clearest neutral-prompt reversal: \textsc{Aligned} is higher at every reportable gate, including strict final success, 7 versus 4.

\mypara{Finding.}
The cross-language check sharpens the claim.
\textsc{Abliterated} is stronger in Java/Vul4J when the task is executable repair and the prompt uses stronger security framing, but that advantage is bounded under neutral prompting.
In neutral cross-language repair, the aligned state is at least competitive and is often more execution-viable.
\Cref{sec:prompt-framing-results} therefore returns from executable substrate effects to the prompt-framing.
Executable repair reinforces the paper's main thesis: safety-state effects are across task depth, prompt framing, and executable substrate.

\section{Prompt-Framing Analysis: Authorization, Security Terminology, and Output Stability}
\label{sec:prompt-framing-results}

The previous sections follow the task-depth progression from diagnostic judgments to executable repair.
This final behavioral experiment returns to prompt-framings.
\Cref{sec:two-factor-overview} showed that coarse prompt frames can move the \textsc{Aligned}/\textsc{Abliterated} gap, but those frames conflate authorization language with security terminology.
The Prompt-Intensity Study separates them: it holds the code sample, task, output schema, and scoring rule fixed, while varying authorization context and lexical intensity.

\subsection{Prompt-Intensity Study Design}
\label{sec:prompt-intensity-design}

This subsection instantiates the bottom-right drill-down in \Cref{fig:method-overview}, where prompt framing is decomposed into authorization context and lexical intensity.
It is the controlled version of the prompt-framing defined in \Cref{sec:method-prompt-framing-conditions}.
We evaluate two tasks from the overview, line localization and CWE attribution, because they expose different kinds of prompt sensitivity.
Localization lets us observe line-set drift, while CWE attribution lets us observe semantic-label drift.
The study uses two contexts, Baseline Context and Authorized Context, crossed with the four lexical levels shown in \Cref{fig:method-overview}: Neutral, Scenario, Tactic/Pattern, and Dense Cyber.

The design is balanced, with only the authorization clause and lexical-intensity wording change.
For each context, the benchmark contains 300 localization selections and 397 CWE selections, evaluated under four lexical levels for both model states.
This gives 2,788 rows per state and 5,576 rows per context.
The design turns prompt wording into a controlled input intervention rather than a post hoc prompt collection.

\subsection{Utility Trajectories Across Context and Lexical Intensity}
\label{sec:prompt-intensity-utility}

\begin{figure*}[!ht]
\centering
\includegraphics[width=\textwidth]{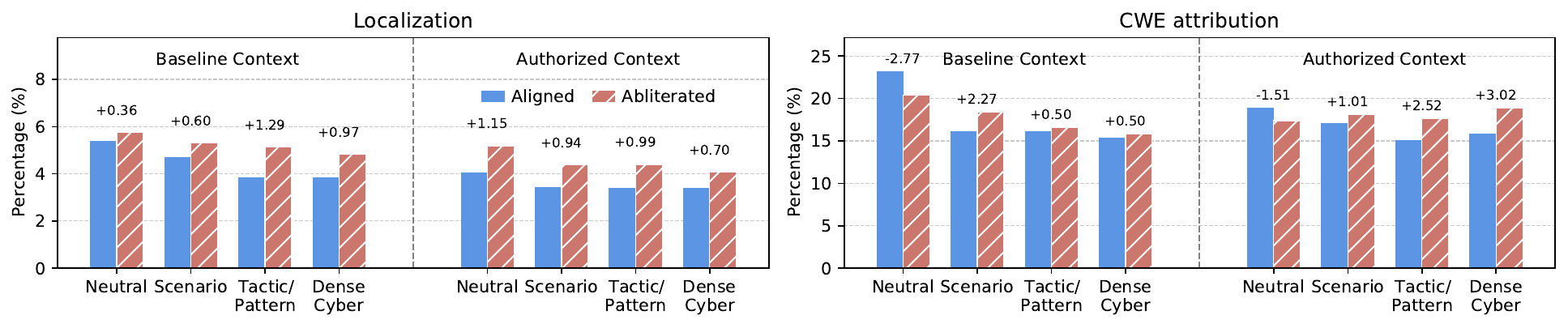}
\vspace{-0.8cm}
\caption{Prompt-intensity utility under Baseline Context and Authorized Context.
The figure reports localization and CWE-attribution utility across Neutral, Scenario, Tactic/Pattern, and Dense Cyber wording for \textsc{Aligned} and \textsc{Abliterated}; the labels above paired bars give \textsc{Abliterated}-minus-\textsc{Aligned} differences in percentage points.}
\label{fig:prompt-intensity-utility}
\vspace{-0.3cm}
\end{figure*}

We first examine how utility changes as prompts move from neutral review wording to denser security terminology.
\Cref{fig:prompt-intensity-utility} reports end-to-end utility for both tasks, both contexts, and all four lexical levels. 
Utility measurement uses the same metric as the overview: line-F1 for localization and exact match for CWE attribution, with unusable responses assigned zero utility.
Positive difference values favor \textsc{Abliterated}; negative values favor \textsc{Aligned}.

\begin{figure*}[!ht]
    \centering
    \includegraphics[width=\textwidth]{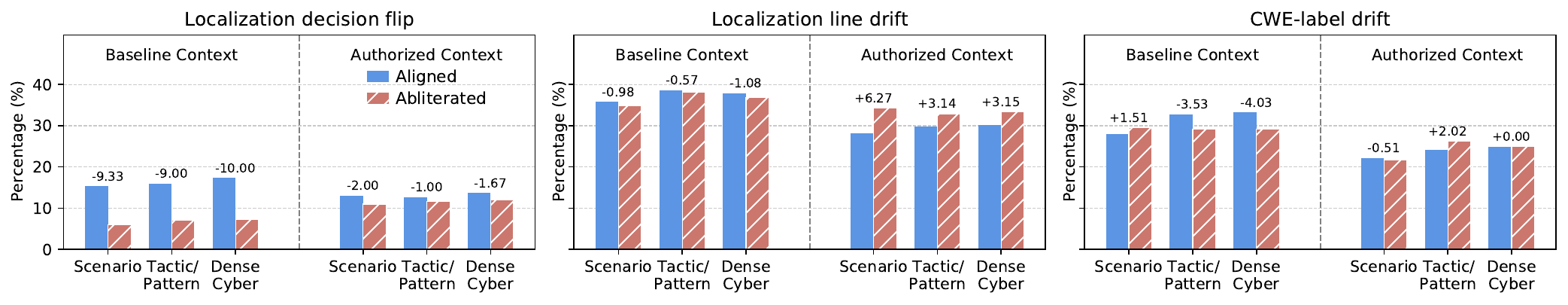}
    \vspace{-0.8cm}
    \caption{Output drift induced by prompt wording relative to Neutral wording.
The figure reports localization decision flips, localization line-set drift, and CWE-label drift under Baseline and Authorized contexts; higher values mean larger behavioral movement from Neutral lexical level within the same context.
The labels above paired bars give \textsc{Abliterated}-minus-\textsc{Aligned} differences in percentage points.}
    \label{fig:prompt-intensity-drift}
\end{figure*}

\Cref{fig:prompt-intensity-utility} shows two prompt-framing patterns.
First, localization consistently favors \textsc{Abliterated} across both contexts and all lexical levels.
The largest Baseline Context gap appears at Tactic/Pattern wording, where \textsc{Abliterated} leads by 1.29 percentage points.
Second, CWE attribution changes direction with wording.
Under Neutral wording, \textsc{Aligned} is higher in both contexts.
Once scenario or cyber-taxonomy terms are introduced, the gap moves toward \textsc{Abliterated}.
This mirrors the coarse-frame result from \Cref{tab:overview-frame-utility}, but now under controlled authorization and lexical-intensity.

The utility trajectories are descriptive.
Across Baseline Context, a linear fit to \textsc{Abliterated}-minus-\textsc{Aligned} differences gives +0.25 percentage points per lexical level for localization and +0.81 for CWE attribution.
\Cref{sec:prompt-intensity-stability} next checks whether these score movements correspond to changed labels or line sets.
The main result is the structured utility trajectory: prompt wording reorients the state gap, with neutral wording preserving the clearest \textsc{Aligned} advantage for CWE and security-explicit wording plus localization moving the relative advantage toward \textsc{Abliterated}.

\subsection{Output Stability Under Prompt Perturbation}
\label{sec:prompt-intensity-stability}

Utility alone understates the prompt-framing effect because two prompts can receive similar scores while changing the concrete answer returned for the same code sample.
We therefore measure output stability against the Neutral lexical level within each context.
For each model state and task instance, we compare the response under Scenario, Tactic/Pattern, or Dense Cyber wording with the response under Neutral wording.
For localization, decision flip reports the percentage of instances whose parsed localization decision changes, and line-set drift reports how much the predicted source-line set changes relative to the Neutral prediction.
For CWE attribution, CWE-label drift reports the percentage of instances whose predicted CWE label changes relative to the Neutral prediction.

\Cref{fig:prompt-intensity-drift} shows that prompt wording changes the answer itself, not only the score.
Under Baseline Context, \textsc{Aligned} has much higher localization decision-flip rates than \textsc{Abliterated}, while both states show substantial line-set drift.
For CWE attribution, both states drift heavily from the Neutral label, with \textsc{Aligned} increasing from 27.96\% to 33.25\% across lexical levels and \textsc{Abliterated} staying near 29\%.
Under Authorized Context, localization decision flips become closer across states, but line-set drift remains large.
The prompt-framing effect is therefore behavioral as well as numerical: benign wording changes can alter the predicted label or line set even when the task and code are unchanged.

\mypara{Finding.}
Together, the utility and drift results close the loop opened by \Cref{sec:two-factor-overview}.
The coarse-frame pattern persists under the auditable authorization-by-intensity design specified in \Cref{sec:method-prompt-framing-conditions}.
Read alongside the task-depth results in \Cref{sec:focused-localization,sec:patch}, the tight claim is that safety-state effects are prompt-contingent: the same benign vulnerability-analysis task can shift in utility and output stability when authorization context and security terminology change.

\section{Conclusion}

This paper studied how LLM safety state affects defender-side utility in vulnerability-analysis workflows. Through matched same-lineage comparisons between aligned instruction-tuned models and refusal-ablated descendants, we show that safety-state effects are context-dependent rather than monotonic. Refusal coverage alone does not explain these effects: both states usually answer, but differ in non-refused answer quality, output stability, localization behavior, and executable repair throughput. The aligned state is often stronger for shallow diagnostic tasks under neutral review wording, whereas the refusal-ablated state becomes more competitive when prompts are security-explicit or tasks require code-grounded and executable assistance. These findings call for cyber-safety evaluations that measure not only refusal and harmful compliance, but also whether legitimate defensive outputs are correct, stable, localized, and actionable enough to survive validation. The goal is not to weaken safeguards, but to design them so they prevent harmful assistance while preserving reliable support for defenders who need precise security language to analyze and repair real vulnerabilities.

\bibliographystyle{IEEEtran}
\bibliography{ccs}

\end{document}